\documentclass[prd,aps,showpacs, superscriptaddress,10pt]{revtex4-1}
\usepackage[utf8]{inputenc}
\usepackage{amsmath}
\usepackage{amssymb}
\usepackage{wasysym}
\usepackage{epsfig}
\usepackage{bbold}

\usepackage{graphicx}
\usepackage{subfigure}

\newcommand{\e}{\mbox{e}}

\begin{document}
 
\title{About the cosmological constant in geometric scalar theory of gravity}

\author{I. C. Jardim}  
\email{jardim@fisica.ufc.br} \affiliation{Departamento de F\'isica,
  Universidade Federal do Cear\'a, Caixa Postal 6030, Campus do Pici,
  60455-760 Fortaleza, Cear\'a, Brazil}

\author{R. R. Landim}  
\email{renan@fisica.ufc.br} \affiliation{Departamento de F\'isica,
  Universidade Federal do Cear\'a, Caixa Postal 6030, Campus do Pici,
  60455-760 Fortaleza, Cear\'a, Brazil}
  
  

\begin{abstract}
 In this paper we study how to include the cosmological constant in geometric scalar theory of gravity (GSG). Firstly we show that the cosmological constant
 could not be modeled by a matter field, unlike in General Relativity. We also show that a spherically symmetric matter distribution, over the de Sitter vacuum,
 does not produce the Kottler solution and no black hole. To circumvent this problem we introduce an coupling term between the scalar field and the vacuum curvature
 in way to provide the Kottler solution. We also apply the original (GSGI) and the modified (GSGII) geometric scalar theory of gravity to the Friedmann-Robertson-Walker cosmology.
 A numerical analysis indicates that GSGII is most sensible to the cosmological constant them GSGI.
\end{abstract}

\maketitle
\section{Introduction}

The special relativity theory postulates that the velocity of light is constant in all coordinate system and produce as a consequence that no signal could travels beyond this speed. This result 
shows that the Newton's gravitational theory wants to be changed. Since the Newtonian theory of gravity was described by a scalar field in Euclidean space, the most simplest way to change
the theory to agree with the special relativity is by a scalar field description in Lorentzian space-time. 
This description was proposed, firstly by Nordström \cite{nordstrom} following by Einstein and Grossmann \cite{Einstein}, and was rapidly discarded by observations. The difficulties of this models
are described, e.g., in Refs. \cite{Deruelle:2011wu, Giulini:2006ry}, and live in Einstein-Grossmann hypothesis that the Minkowski background is observable. Recently a new proposal to describe the gravity
by a scalar field was made, called geometric scalar theory of gravity (GSG) \cite{Novello:2012wr}. The key of this new theory is that this scalar field produces an observed non-trivial geometry in presence of matter, while the 
Minkowski in unobserved, violating the Einstein-Grossmann hypothesis. In this way the gravity is related to the geometry of space-time, like in General Relativity, and it is free of all objections made
to scalar description of gravity ( see, for instance \cite{Feymann}). As showed in Ref. \cite{Novello:2012wr} this description agree with all gravitational tests, like the solar system experiments. The cosmology of this model was study
in a recent paper and shows the possibility of a bounce without an exotic cosmic fluid \cite{Bittencourt:2014oua}. 

The cosmological constant was introduced by Einstein to agree with a static universe model. Despite this model has been replaced by a dynamical one, the Friedmann-Robertson-Walker model,
the most recent observations of the expansion rate of the Universe indicates the existence of a non-vanish cosmological constant \cite{Perlmutter:1998np, Riess:1998cb}. 
This discovery won the Nobel Prize in 2011 for their discoverers. In a mathematical point of view this kind of term was two interesting properties, the first one is that they are the only term that is compatible with covariant energy-momentum tensor conservation;
The second point is that this cover all space-time with non-vanish constant curvature. The positive one (the anti-de Sitter space-time) is generated by a negative cosmological constant and the negative one (the de Sitter
space-time) by a positive cosmological constant \cite{hawkingEllis}.
In physics, beyond the cosmological implication, some results in string theory needs a non-vanish constant curvature. This is the case of the correspondence of conformal field theory and
the anti-de Sitter space-time ( AdS/CFT correspondence) \cite{Maldacena:1997re}, which is the subject of uncountable recent works. An another area which the cosmological constant is essential is in brane scenarios
with extended extra dimensions, like the Randall-Sundrum one \cite{Randall:1999vf, Jardim:2011gg, Jardim:2013jy}.

In this paper we use two approaches to include the cosmological constant in GSG. The first one is modeling the cosmological constant as a matter field living in the Minkowski background.
This is the first way due the scalar field in GSG lives on flat space-time, so the change in effective metric is given by matter fields. Since the metric is over-determined the first step is show
that a scalar field is able to produce the de Sitter (anti-de Sitter) space-time, the second step is show that it can be made by an acceptable energy density distribution.
The second approach used in this work to include the cosmological constant is changing the vacuum state of the theory. In this case the background that the scalar field lives is not the Minkowski
space-time, but the de Sitter/Anti-de Sitter one. After this we  show that the solution for a spherically symmetric matter distribution over this background generates an acceptable solution, but not the standard
Kottler solution. Since the found solution does not produce a black hole we modify the field equation to be compatible with the black hole physics. We called this modified theory as
geometric scalar theory of gravity type II (GSGII). Since the cosmological constant play an important hole in cosmology we apply the both type of GSG in Friedmann-Robertson-Walker metric.

This work was organized as following: In Sec. II we make a short review of geometric scalar theory of gravity. We present the fundamental hypothesis and the field equation of this theory of gravity
that we will use in following sections. In Sec. III we model the cosmological constant as a matter field living in a Minkowski background. In Sec. IV we insert the cosmological constant as the vacuum
state and compute the solution for a spherically symmetric matter in this background. In Sec. V we modify the field equation to support the Kottler solution as the solution for a spherically symmetric
matter in non-vanish constant curvature background. In Sec. VI we apply the booth versions of GSG for cosmology. We illustrate the behavior of scale constant in booth versions and comment the differences.
Finally we discuss the conclusions and perspectives.
\section{Review of Geometric scalar theory of gravity}
In this section we will make a short review of geometric scalar theory of gravity (GSG) based in Ref. \cite{Novello:2012wr}. 
The fundamental hypothesis of GSG is that the gravitational interaction is mediated by the scalar field $\Phi$.
All forms of matter and energy interact with $\Phi$ only through the metric $q_{\mu\nu}$ and its derivatives in a covariant way.
The contravariant metric tensor $q^{\mu\nu}$, following the Ref. \cite{Novello:2011sh}, is given by
\begin{equation}\label{qcontra}
 q^{\mu\nu} = \alpha\eta^{\mu\nu} +\frac{\beta}{w}\partial^{\mu}\Phi\partial^{\nu}\Phi,
\end{equation}
where $\eta^{\mu\nu}$ is the Minkowski metric, $\partial^{\mu}\Phi \equiv \eta^{\mu\nu}\partial_{\nu}\Phi$, $w \equiv \eta^{\mu\nu}\partial_{\mu}\Phi\partial_{\nu}\Phi$; $\alpha$ and $\beta$ are scalar functions of $\Phi$.
The authors in \cite{Novello:2012wr} search for the functions $\alpha$ and $\beta$ in way to obtain the following field equation 
\begin{equation}
 \Box\Phi = 0,
\end{equation}
in absence of matter, where $\Box$ is the Laplace-Beltrami operator relative to the metric $q_{\mu\nu}$, i.e.,
\begin{equation}
\Box\Phi = \frac{1}{\sqrt{-q}}\partial_{\mu}\left[\sqrt{-q}q^{\mu\nu}\partial_{\nu}\Phi\right]. 
\end{equation}

Extrapolating the Newtonian limit, the authors in  \cite{Novello:2012wr} write
$\alpha = \mbox{e}^{-2\Phi}$
and, to obtain the Schwarzschild metric, they fix
\begin{equation}
Z \equiv \beta +\alpha = \frac{(\alpha -3)^{2}}{4}.
\end{equation}
The geometric scalar theory of gravity could be derived from the action formalism. In the unobservable auxiliary Minkowski background the gravitational action is given by
\begin{equation}
 S = \int\sqrt{-\eta}d^{4}x V(\Phi)w,
\end{equation}
where $V(\Phi)$ is a potential which relates with $\alpha$ by
\begin{equation}
V(\Phi) = \frac{(\alpha-3)^{2}}{4\alpha^{3}}.
\end{equation}
Taking the variation in action , we obtain
\begin{equation}
 \delta S = -2\int\sqrt{-q}d^{4}x\sqrt{V}\Box\Phi\delta\Phi.
\end{equation}
In the presence of matter, taking the variation of matter action with respect to $\Phi$ we obtain the field equation
\begin{equation}\label{GSGI}
\sqrt{V}\Square\Phi = \kappa\chi.
\end{equation}
The matter content is represented by the scalar,
\begin{equation}\label{chi}
 \chi = \frac{1}{2}\left[\frac{3\e^{2\Phi} +1}{3\e^{2\Phi} -1}E -T -C^{\mu}_{;\mu}\right];
\end{equation}
where
\begin{equation}
T \equiv q_{\mu\nu}T^{\mu\nu},\;\;\;\; E = \frac{T^{\mu\nu}\partial_{\mu}\Phi\partial_{\nu}\Phi}{\Omega},\;\;\;\; C^{\mu} = \frac{\beta}{\alpha\Omega}\left(T^{\mu\nu} -Eq^{\mu\nu}\right)\partial_{\nu}\Phi,
\end{equation}
 $\Omega = Zw$ and $\kappa = 8\pi G$. As showed in \cite{Novello:2012wr}, for a static perfect fluid in a homogeneous time-dependent space-time, the gravitational source (\ref{chi}) can be decomposed as
\begin{equation}\label{dec1}
 \chi = -\frac{1}{2}\left[\frac{2\alpha}{\alpha -3}\rho -3p\right],
\end{equation}
where $\rho$ is the energy density and $p$ is the pressure. 

\section{Cosmological Constant as a matter field}
In this section we will try to model the cosmological constant by a matter field. Beginning with the auxiliary Minkowski space-time in spherical coordinates
\begin{equation}
 ds_{0}^{2} = dt^{2} -dR^{2} -R^{2}d\Omega^{2},
\end{equation}
and changing the radial coordinate to $R = r\sqrt{\alpha(r)}$, we can use the relation (\ref{qcontra}) to obtain the gravitational metric
\begin{eqnarray}
q_{00} &=& \frac{1}{\alpha}, \label{q00}
\\q_{11} &=& -\frac{\alpha}{\alpha +\beta}\left[\frac{1}{2\alpha}\frac{d\alpha}{dr}r +1\right]^{2}, \label{q11}
\\ q_{22} &=& -r^{2} \;\;\;\;\;\;\mbox{and}\;\;\;\;\;\;\ q_{33} = -r^{2}\sin^{2}\theta. \label{qang}
\end{eqnarray}
Similar to (\ref{dec1}), for a static perfect fluid in a static and spherically symmetric space-time, the gravitational source (\ref{chi}), can be decomposed as
\begin{equation}
 \chi = \frac{3-2\alpha}{3-\alpha}p -\frac{1}{2}\rho.
\end{equation}
The metric elements, (\ref{q00})-(\ref{qang}), in field equation (\ref{GSGI}), with the above source, provides  
\begin{equation}\label{eqamatt}
 \frac{(\alpha -3)^{2}}{2\alpha^{3/2}r^{2}}\left[\frac{1}{2\alpha}\frac{d\alpha}{dr}r +1\right]^{-1}\frac{d}{dr}\left[ r^{2}\frac{|\alpha -3|}{4\alpha^{2}}\left[\frac{1}{2\alpha}\frac{d\alpha}{dr}r +1\right]^{-1}\frac{d\alpha}{dr}\right] 
=  \kappa\frac{3(2\sigma-1) +(1-4\sigma)\alpha}{\alpha -3}\rho,
\end{equation}
where we have used the state equation $p = \sigma\rho$. Since is not possible to solve the above equation analytically, we can check if the de Sitter metric is a solution. Like the metric is over-determined, first
we can observe that, if
\begin{equation}\label{solamatt}
 q_{00} = \alpha^{-1} =  (1-\lambda r^{2}),
\end{equation}
the equation (\ref{q11}) provides, for the radial component of gravitational metric tensor,
\begin{equation}\label{q11matt}
 q_{11} = -\frac{ (1-\lambda r^{2})^{-1}}{(1-3\lambda r^{2}/2)^{2}},
\end{equation}
that is not the de Sitter solution, i.e., a solution with constant nonzero curvature. The gravitational metric component, (\ref{q11matt}), inserts a new singularity at 
$r = (3\lambda/2)^{-1/2}$, near than the de Sitter horizon. Unlike the Schwarzschild and de Sitter, this new singularity can not be removed, as we can observe by the determinant of the geometric metric tensor.  
Indeed the only solution that satisfies the condition
\begin{equation}\label{cond}
 q_{11} = -q_{00}^{-1}.
\end{equation}
is the Schwarzschild solution, as obtained in \cite{Novello:2012wr}. Despite the de Sitter metric could not be found by a matter distribution we will compute
the energy density to produce the metric (\ref{q11matt}) for completeness. Replacing the solution (\ref{solamatt}) in field equation (\ref{eqamatt}) we find the energy density of the matter
\begin{equation}
 k\rho = 3\lambda\frac{(1-3\lambda r^{2}/2)^{3}(2-5\lambda r^{2})(1-\lambda r^{2})^{1/2}}{(\sigma -1) -3(\sigma - 1/2)\lambda r^{2}}.
\end{equation}
If we impose that at the origin the density is positive, $\sigma >1$, the energy density diverges at $r = (3\lambda(\sigma-1/2)/(\sigma -1))^{-1/2}$. To avoid this behavior we can fix $\sigma = -2$ or $\sigma = 0$, and obtain
\begin{equation}
 k\rho = \left\lbrace\begin{matrix}&-2\lambda(1-3\lambda r^{2}/2)^{3}(1-\lambda r^{2})^{1/2}, &\mbox{for}& \sigma = -2,\\ & 
 
 - 3\lambda(1-3\lambda r^{2}/2)^{2}(2-5\lambda r^{2})(1-\lambda r^{2})^{1/2}, &\mbox{for}& \sigma = 0.\end{matrix}\right.
\end{equation}
In the first case the energy density is always negative for a positive lambda, in opposition of in the de Sitter space-time in GR. In the second case the energy density is negative
until $ r = (5\lambda/2)^ {1/2}$, changing the signal after this point.
\\
This result shows that the de Sitter solution is not compatible with the field equations for a spherically symmetric matter, i.e.,
unlike the general relativity theory the de Sitter space-time is not the solution to a matter which has

\begin{equation}
 T_{\mu\nu} \propto q_{\mu\nu}.
\end{equation}

What leads to the conclusion that the geometric-scalar theory of gravitation can decide if the cosmological constant is a matter field or geometry vacuum.
To keep the GSG theory compatible with
de Sitter space we have to postulate that the vacuum of the theory is not the Minkowski space-time but the de Sitter one. 
This behavior could be explained by the fact that the topology of the de Sitter space-time differ than the Minkowski one. Is suitable to think that the scalar field could not change the topology of the background.
Then the background  which the field $ \Phi $ permeates and which the gravitational 
metric $q_{\mu \nu} $ should be expanded is the de Sitter space-time. This vacuum state must provide consistent solutions in presence of matter, to verify this we will compute the Schwarzschild de Sitter solution ( Kottler solution) in next section.

\section{Spherically symmetric matter distribution in de Sitter background}
In Ref. \cite{Novello:2012wr} the authors found the Schwarzschild solution for a symmetric matter distribution, since they consider the vacuum is the Minkowski space-time. As shown in the previous section the de Sitter space-time can not be obtained
from a spherically symmetric distribution of matter, having to be implanted in the vacuum state of the theory. Over this vacuum state a localized spherically symmetric distribution of matter 
must to produce a solution similar to Kottler solution. To verify this we will start from the de Sitter vacuum written in spherical coordinates
\begin{equation}
 ds_{0}^{2} =  (1-\lambda R^{2})dt^{2} - (1-\lambda R^{2})^{-1}dR^{2} -R^{2}d\Omega^{2},
\end{equation}
using the same procedure used in previous section, i.e., change to radial coordinate $R = r\sqrt{\alpha(r)}$ and use the eq. (\ref{qcontra}), we can find the gravitational metric elements
\begin{eqnarray}
 q_{00} &=& \frac{1-\lambda\alpha r^{2}}{\alpha}, \label{1q00}
\\q_{11} &=& -\frac{\alpha(1-\lambda\alpha r^{2})^{-1}}{\alpha +\beta}\left[\frac{1}{2\alpha}\frac{d\alpha}{dr}r +1\right]^{2}, \label{1q11}
\\ q_{22} &=& -r^{2} \;\;\;\;\;\;\mbox{and}\;\;\;\;\;\;\ q_{33} = -r^{2}\sin^{2}\theta. \label{1qang}
\end{eqnarray}
The field equation (\ref{GSGI}), out of matter distribution, with the above gravitational metric provides
\begin{equation}\label{eqkott}
 \left[r(3 -\alpha)( 1-\lambda\alpha r^{2})+2\alpha\Phi_{0}\right]\frac{d\alpha}{dr} +\frac{4\alpha^{2}}{r}\Phi_{0} = 0,
\end{equation}
where $\Phi_{0}$ is an integration constant. Firstly, we will test if the Kottler solution satisfy the field equation. Imposing this solution in time component of gravitational metric, (\ref{1q00}), we find that  
\begin{equation}\label{ak}
 \alpha  =  \left(1 -\frac{2GM}{r}\right)^{-1}.
\end{equation}
As commented in previous section, this solution satisfy the condition (\ref{cond}), i.e., the radial component of gravitational metric, (\ref{1q11}), is also the  Kottler solution. But replacing
(\ref{ak}) in field equation (\ref{eqkott}) we obtain the fixation
\begin{equation}
 \Phi_{0} = GM( 1-\lambda\alpha r^{2})
\end{equation}
which is not compatible with the fact that $\Phi_{0}$ is a constant. So, we conclude that the Kottler solution does not satisfy the field equation.
\\ Due the nonlinearities of field equation (\ref{eqkott}) it is not possible to obtain an analytic solution. But we can prove that this equation does not produce a solution with a similar behavior of the Schwarzschild solution.  
To facilitate the analysis of this equation we will rewrite it in terms of $ q_{00} = q(r)$. In this variable the equation becomes
\begin{equation}
 \frac{dq}{dr} = 2\frac{ -\lambda r^{2}\left(3(q +\lambda r^{2}) -1\right)q +2\Phi_{0}\frac{q}{r}(q +\lambda r^{2})}{\left(3(q +\lambda r^{2}) -1\right)qr+2\Phi_{0}(q +\lambda r^{2})},
\end{equation}
and the radial component can be written as
\begin{equation}
 q_{11} =-\frac{4qr^{2} (q + \lambda r^{2})^{2}}{[\left(3(q +\lambda r^{2}) -1\right)qr+2\Phi_{0}(q +\lambda r^{2})]^{2}}.
\end{equation}

To have a acceptable behavior the solution must to provides at least one horizon for $\lambda > 0$, since we expect that far away the matter distribution the solution approach to de Sitter solution and provides a horizon. To produce a horizon is necessary that 
the null curves approaches asymptotically to a  finite point $r = r_{h}$, i.e., 
\begin{equation}
\lim_{r\to r_{h}}\left(\frac{dt}{dr}\right)^{2} = \lim_{r\to r_{h}} \frac{q_{11}(r)}{q(r)} = \infty.
\end{equation}
To satisfy the above limit is necessary that, or $q \to 0$ or $q_{11}\to\infty$ when $r\to r_{h}$. The first possibility does not provides a physical solution due, as we can show,  when $q \to0$ all derivatives of $q$ vanishes,
indicating that this can not be a finite point-. To perform the second possibility
is necessary that in $r = r_{h}$ the denominator of $q_{11}$ vanishes, i.e.,
\begin{equation}
 \left(3(q_{h} +\lambda r_{h}^{2}) -1\right)q_{h}r_{h}+2\Phi_{0}(q_{h} +\lambda r_{h}^{2}) = 0,
\end{equation}
were $q_{h} \equiv q(r_{h})$, which in nonzero and, by above expression, has the value
\begin{equation}
 q_{h} = -\frac{[2\Phi_{0} + r_{h}(3\lambda r_{h}^{2}-1)]\pm \sqrt{[2\Phi_{0} + r_{h}(3\lambda r_{h}^{2}-1)]^{2} -24\Phi_{0}\lambda r_{h}^{3}}}{6r_{h}}.
\end{equation}
The condition that the metric must to be positive and real, provides the following constraint to integration constant
\begin{equation}
\Phi_{0} < \left\lbrace\begin{matrix}                     
-\sqrt{3\lambda}r_{h}^{2} + (r_{h} + 3 r_{h}^{3} \lambda)/2 &,\mbox{if}& 0 < \lambda < (3 r_{h}^{2})^{-1},\\
 (r_{h} - 3 r_{h}^{3} \lambda)/2 &,\mbox{if}& \lambda \geq (3 r_{h}^{2})^{-1}.
\end{matrix}
 \right.
\end{equation}
\begin{figure}[!h]
\centering
\subfigure[]{
\includegraphics[scale=0.3]{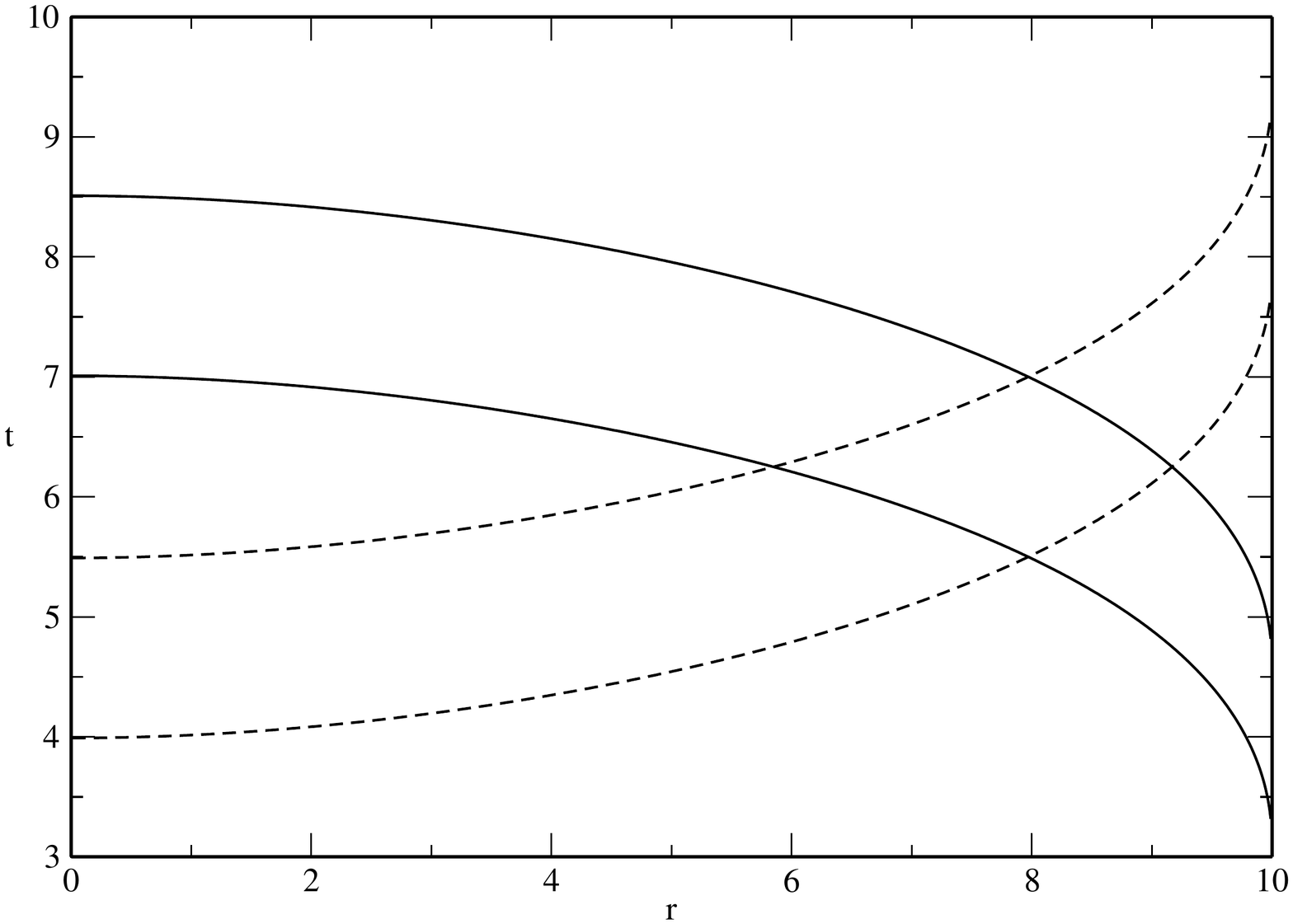}
 \label{fig:tcond1}
}
\subfigure[]{
\includegraphics[scale=0.3]{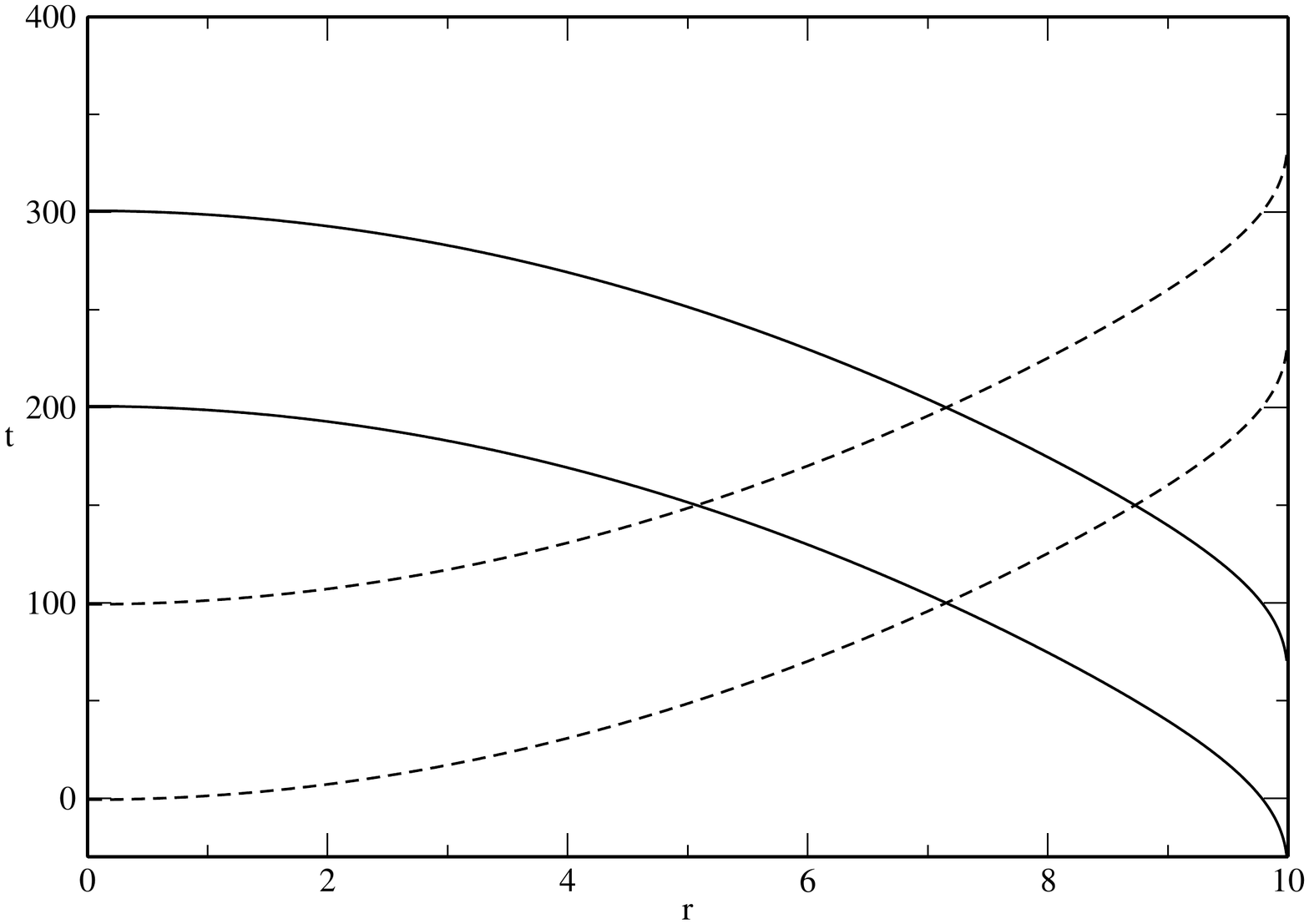}
 \label{fig:tcond2}
}
\caption{Causal structure of space-time (a) for $ \lambda < (3 r_{h}^{2})^{-1}$ and (b) for $ \lambda \geq (3 r_{h}^{2})^{-1}$. The continuous lines indicates the in-going light rays and the dashed lines indicates the outgoing light rays. }
\end{figure}

The behavior of light-like surfaces are illustrated in fig. \ref{fig:tcond1} for  $0 < \lambda < (3 r_{h}^{2})^{-1}$ and in fig. \ref{fig:tcond2} for  $\lambda \geq (3 r_{h}^{2})^{-1}$. As we can see in booth cases that
are no black holes, i.e., no horizon are generated by matter distribution. If we want to keep the black holes in geometric scalar theory of gravity with cosmological constant we must to change
the field equation. This is what we will do in next section.

\section{Geometric Scalar Theory of Gravity type II}
In previous section, we show that a spherical distribution of matter in de Sitter vacuum does not produce a black hole. In general relativity type II the Einstein equation produce, for this kind of matter
distribution, the Kottler solution, i.e., Schwarzschild de Sitter solution. As we proved previously, the field equation of geometric scalar theory of gravity is not satisfied by Kottler solution. 
In this section we will search a new field equation for a scalar theory of gravity which is satisfied by this solution. 
\\As computed in previous section, to obtain the Kottler metric starting from the de Sitter vacuum, the field $\alpha$ is given by (\ref{ak}) and the metric elements is given by (\ref{1q00})-(\ref{1qang}). 
Computing the Laplace-Beltrami operator with this metric elements we obtain that
\begin{equation}
 \Box\Phi = \frac{2\lambda GM}{r}\left(1-\frac{3GM}{r}\right)\left(1-\frac{2GM}{r}\right)^{-2}.
\end{equation}
Replacing $GM/r$ by $\alpha$, using the eq. (\ref{ak}), we can rewrite the above result in the way 
\begin{equation}
 \Box\Phi  +\frac{\lambda}{2}(\alpha-1)(\alpha-3) = 0.
\end{equation}
This is the field equation of scalar theory of gravity type II (GSGII) in absence of matter. The new term can be understood as a coupling of scalar field with the curvature of vacuum state. This term have two special points, $\alpha = 1$ and $\alpha = 3$. The first point is the vacuum solution, i.e.,
the existence of this zero ensure that the de Sitter is the vacuum solution. The second point,  $\alpha = 3$, are related with the zero of the potential $V(\Phi)$. In this way no new special points are introduced 
by the new term.
In presence of matter the field equation of scalar theory of gravity type II can be written as
\begin{equation}\label{GSGII}
\sqrt{V}\left[ \Box\Phi  +\frac{\lambda}{2}(\alpha-1)(\alpha-3)\right] = \kappa\chi.
\end{equation}
This is the field equation that produces the Kottler solution by a spherically symmetric mass distribution and agree with all solar tests and black holes physics. 

\section{Cosmology}
\begin{figure}[!h]
\centering
\subfigure[]{
\includegraphics[scale=0.3]{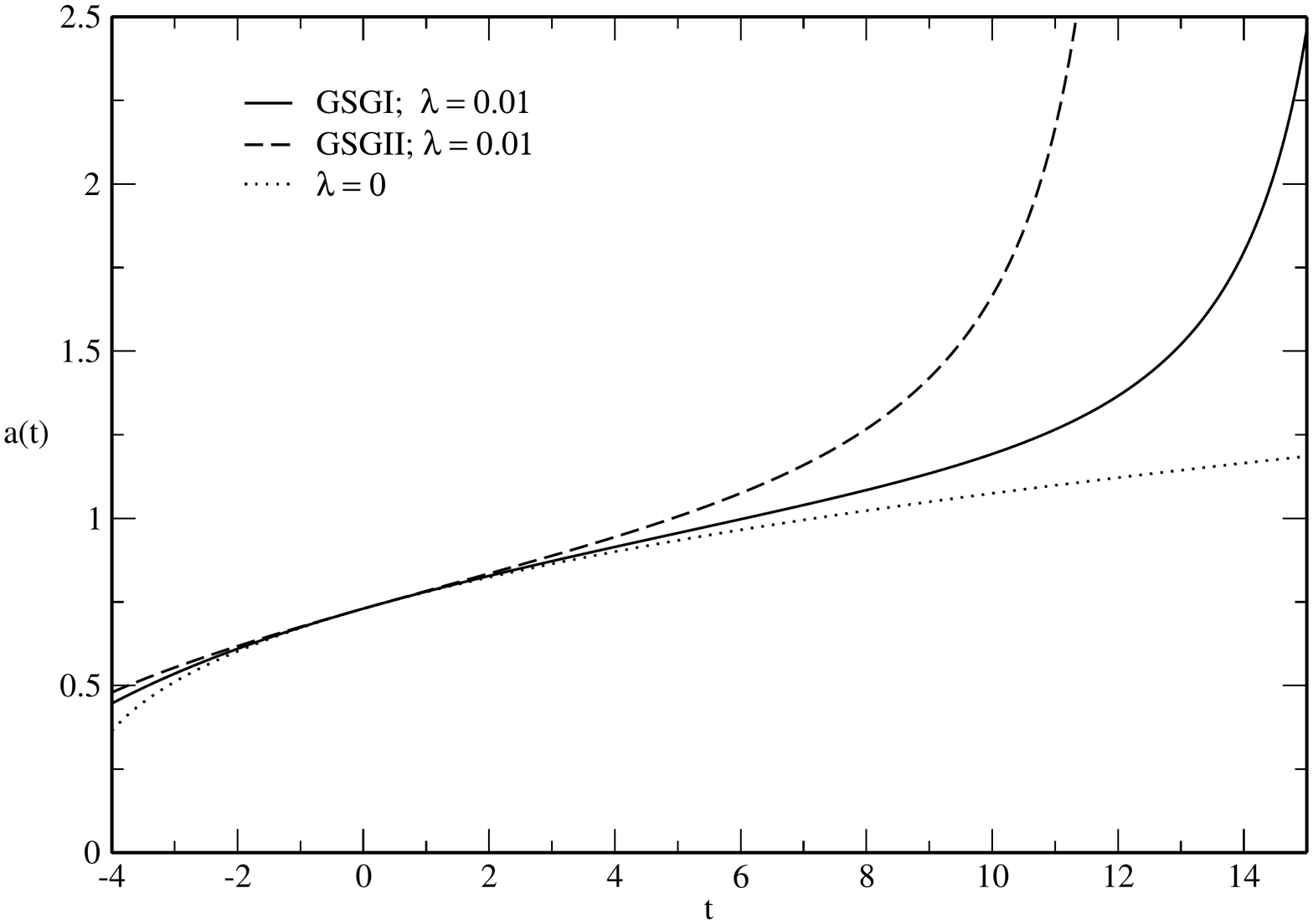}
 \label{fig:aIIxaI}
}
\subfigure[]{
\includegraphics[scale=0.3]{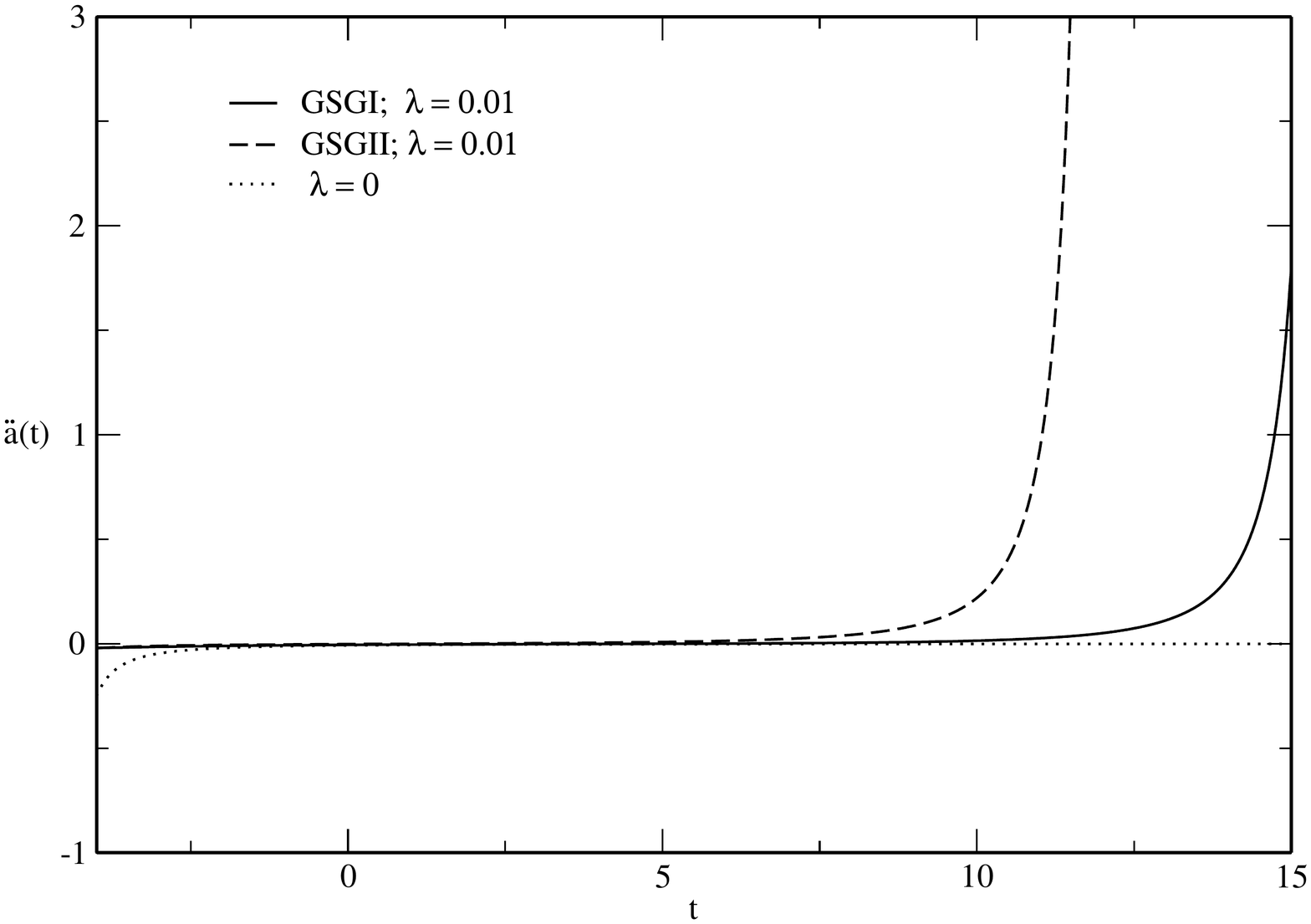}
 \label{fig:acIIxacI}
}
\caption{The comparison of behavior of scale factor in both geometric scalar theories of gravity. In this simulation we use the initial conditions $a(0) = 0.73$, $ \dot{a}(0)/a(0) = 1/13.7$ and $T(0) = 0$.}
\end{figure}

The de Sitter space-time produces several consequences in large scale, due this, it have a fundamental importance in cosmology. When the general relativity theory is applied to the cosmos, its needs a positive cosmological constant to
fit the accelerated expansion ratio. In previous sections we show two ways to introduce the cosmological constant in geometric scalar theory of gravity.
In this section we will analyze the influence of this two approach in cosmology. 
\\ Beginning with the de Sitter unobserved vacuum in the form
\begin{equation}
 ds^{2}_{0} = -dT^{2} +\e^{2T/l}(dR^{2} +R^{2}d\Omega^{2}),
\end{equation}
where $l = \lambda^{-1/2}$, we can use the eq. (\ref{qcontra}) to write the gravitational metric for $\Phi = \Phi(T)$,
\begin{equation}
ds^{2} = -\frac{dT^{2}}{\alpha +\beta} +\frac{\e^{2T/l}}{\alpha}(dR^{2} +R^{2}d\Omega^{2})
\end{equation}
To write in Friedmann-Robertson-Walker form we define the comoving time and the scale factor as
\begin{equation}
 dt = \frac{dT}{\sqrt{\alpha +\beta}} \;\;\;\;\mbox{and}\;\;\; a^{2}(t) = \frac{\e^{2T/l}}{\alpha},
\end{equation}
respectively. The Laplace-Beltrami operator in Friedmann-Robertson-Walker metric can be written as
\begin{eqnarray}
 \Box\Phi &=& \frac{1}{2l}\left[-\frac{\e^{2T(t)/l}}{a^{4}(t)l}\left(3a^{2}(t)-\e^{2T(t)/l}\right) +\left(9a^{2}(t)-\e^{2T(t)/l}\right)\frac{\dot{a}(t)}{a^{3}(t)} \right] -\left[2\left(\frac{\dot{a}(t)}{a(t)}\right)^{2} +\frac{\ddot{a}(t)}{a(t)}\right],
\end{eqnarray}
where the dot indicates derivatives with respect to comoving time, $t$. In the above operator the functions $a(t)$ and $T(t)$ are coupled, to relate them we must to use the definition of the comoving time
\begin{equation}\label{Ttat}
  dt = \frac{2a^{2}(t)}{|\e^{2T/l} -3a^{2}(t)|} dT
\end{equation}
In  geometric scalar theory of gravity type I the equation of motion, eq. (\ref{GSGI}), for a perfect fluid with $p = \sigma\rho$ is given by
\begin{equation}
\sqrt{V}\left[2\left(\frac{\dot{a}}{a}\right)^{2} +\frac{\ddot{a}}{a}\right] -\frac{\sqrt{V}}{2l}\left[-\frac{\e^{2T/l}}{a^{4}l}\left(3a^{2}-\e^{2T/l}\right) +\left(9a^{2}-\e^{2T/l}\right)\frac{\dot{a}}{a^{3}} \right] = \frac{\kappa}{2}\rho_{0}\frac{(2 -3\sigma)\e^{2T/l} +9\sigma a^{2}}{3a^{2} -\e^{2T/l}}a^{-3(1+\sigma)}
\end{equation}
This equation couples $a(t)$ and $T(t)$, and must to be decoupled using (\ref{Ttat}). This procedure can not be made analytically and need a numerical analysis to take conclusions.
In geometric scalar theory of gravity type II the equation of motion, eq. (\ref{GSGII}), for a perfect fluid with $p = \sigma\rho$ is
\begin{equation}
\sqrt{V}\left[2\left(\frac{\dot{a}}{a}\right)^{2} +\frac{\ddot{a}}{a}\right] -\frac{\sqrt{V}}{2l}\left[- \frac{3a^2-\e^{2T/l}}{a^2 l} +(9a^{2}-\e^{2T/l})\frac{\dot{a}}{a^{3}} \right] = \frac{\kappa}{2}\rho_{0}\frac{(2 -3\sigma)\e^{2T/l} +9\sigma a^{2}}{3a^{2} -\e^{2T/l}}a^{-3(1+\sigma)}
\end{equation}
Similar to GSGI case, the above equation must to be decoupled  using (\ref{Ttat}) and need a numerical analysis. To illustrate the behavior of both theories we plot the scale factor in fig. \ref{fig:aIIxaI}
and the cosmic acceleration in fig. \ref{fig:acIIxacI}. Due the high influence of matter density the both graphics was computed in absence of matter, just to illustrate the influence of cosmological constant. 
The both figures indicates that the cosmological constant increases the scale factor quickly in GSGII than in GSGI. The fig. \ref{fig:acIIxacI} indicates that the geometric scalar theory of gravity  
needs a cosmological constant to produce an accelerated universe.

\section{Conclusion}
In this paper we study how the cosmological constant could be inserted in geometric scalar theory of gravity. First of all we show that the GSG could not support a nonvanish constant curvature space-time. Them we also show that the cosmological constant could not be a matter field
with a physically acceptable energy density. This result break the freedom of General Relativity, where the cosmological constant can be understood as a matter field also a non-trivial vacuum geometry.
We show that in geometric scalar theory of gravitation the only way to insert the cosmological constant is as a vacuum state, i.e., the vacuum state is not Minkowski, but de Sitter/ anti de Sitter space-time.
 
In sec. IV we show that the GSG with nonzero constant curvature vacuum state does not produce the Kottler solution. Using numerical analysis we conclude that no black holes are generated 
by a spherically symmetric distribution of matter. The degrees of freedom permits that this model fits the solar experiments, but not all characteristics of Schwarzschild solution. This
scenario indicates that the final state of star evolution can not be a black hole, in opposition of current theories. 
To solve the non-existence of black holes we proposes a modification in the field equation which contains an interaction term between the scalar gravitational field and the vacuum state. This modification
restore the Kottler solution and fit the solar experiments and the standard star evolution. We called this modified theory as geometric scalar theory of gravitation type II (GSGII). 
All results obtained support the limit of the vanish vacuum curvature, providing a generalization of the theory presented in Ref. \cite{Novello:2012wr}

Finally we apply the both theories of gravitation in cosmological scenario. We obtain the coupled set of field equation which governs the evolution of scale factor in both cases.
Numerically we show that the cosmological constant accelerate the universe quickly in GSGII than in GSGI. 

\section*{Acknowledgments}

The authors thanks to Junior Diniz Toniato for a critical reading. We acknowledge the financial support provided by Funda\c c\~ao Cearense de Apoio ao Desenvolvimento Cient\'\i fico e Tecnol\'ogico (FUNCAP), the Conselho Nacional de 
Desenvolvimento Cient\'\i fico e Tecnol\'ogico (CNPq) and FUNCAP/CNPq/PRONEX.

\end{document}